\documentclass[reprint,amsmath,amssymb,aps,prl,amssymb,superscriptaddress,notitlepage,floatfix]{revtex4-1}
\usepackage[utf8]{inputenc}
\usepackage{color}
\usepackage{amsmath, braket}
\usepackage{mathtools}
\usepackage{bm}
\usepackage{amsfonts}
\usepackage{dcolumn}
\usepackage{natbib}
\usepackage{bbold}
\usepackage{soul}
\usepackage[dvipsnames]{xcolor}
\usepackage{amssymb}
\usepackage[breaklinks=true,colorlinks,citecolor=blue,linkcolor=blue,urlcolor=blue]{hyperref}
\usepackage{graphicx}
\usepackage{appendix}

\usepackage{mathtools}

\usepackage{pifont}  
\newcommand{\cmark}{\ding{51}}%
\newcommand{\xmark}{\ding{55}}%

\usepackage{soul}

\begin{document}
\title{Correlation-Driven Orbital Order Realizes 2D Metallic Altermagnetism}
\author{Nirmalya Jana} 
\email{nirmalyaj20@iitk.ac.in}
\affiliation{Department of Physics, Indian Institute of Technology Kanpur, Kanpur-208016, India}
\author{Atasi Chakraborty} 
\email{atasi.chakraborty@uni-mainz.de}
\affiliation{Institut f\"{u}r Physik, Johannes Gutenberg Universit\"{a}t Mainz, D-55099 Mainz, Germany}
\author{Anamitra Mukherjee}
\email{anamitra@niser.ac.in}
\affiliation{School of Physical Sciences, National Institute of Science Education and Research, a CI of Homi Bhabha National Institute, Jatni 752050, India} 
\author{Amit Agarwal} 
\email{amitag@iitk.ac.in}
\affiliation{Department of Physics, Indian Institute of Technology Kanpur, Kanpur-208016, India}

\begin{abstract}
Two-dimensional metallic altermagnets are rare, and no correlated 2D material has been established to host large nonrelativistic spin splitting. Here we show that spontaneous orbital order, driven by electronic correlations and Fermi surface nesting, provides a general microscopic route to two-dimensional metallic altermagnetism. Antiferro-orbital ordering between the d$_{xz}$ and d$_{yz}$ orbitals breaks the equivalence of magnetic sublattices with opposite spins and generates a symmetry-enforced altermagnetic spin texture. As a concrete realization, we identify monolayer YbMn$_2$Ge$_2$ as a stable correlated metallic altermagnet exhibiting giant nonrelativistic spin splitting of order 1 eV. The resulting phase supports an exceptionally large and gate-tunable transverse spin conductivity. These results establish correlation-driven orbital order as a robust and general mechanism for designing correlated altermagnets with large spin splitting. 
\end{abstract}

\maketitle 

\textit{Introduction:-} 
The lifting of Kramers spin degeneracy in collinear antiferromagnets with zero net magnetization has led to the identification of a distinct magnetic phase known as \emph{altermagnetism}~\cite{beyond_fm_and_afm_prx_smejkal_22,
research_landscape_of_altermagnets_smejkal_22}. In conventional antiferromagnets, sublattice-connecting symmetries such as $\mathcal{PT}$ or $t_{1/2}\mathcal{T}$ protect spin-degenerate bands despite opposite local moments~\cite{smejkal_NRM_AHA,rmp_baltz_afm_spintronics_2018}. In altermagnets, the breaking of these symmetries removes Kramers degeneracy even in collinear spin configurations, producing momentum-dependent spin splitting without generating a net magnetization~\cite{pt_tt_broken_AM_zhou_2025,broken_kramers_in_AM_MnTe_lee_2024,AM_lifting_of_Kramers_degeneracy_Krempasky_2024}. Recent symmetry-based classifications and minimal model constructions have enabled the targeted search for altermagnetic materials~\cite{smejkal_symmetry_based_altermagnetism_prx_22,
research_landscape_of_altermagnets_smejkal_22,
2D_AM_by_symmetry_Zeng_24,
AM_model_by_symmetry_roig_24,
hayami_design_of_spin_split_bands_in_antiferromagnets_20}. Several candidate systems have now been experimentally identified~\cite{AM_lifting_of_Kramers_degeneracy_Krempasky_2024,
search_for_metallic_AM_wan_2025,
2D_ferroelectric_AM_model_to_material_zhu_2025}, significantly advancing the field. In most of these materials, the anisotropic spin-dependent hopping responsible for altermagnetism is directly imposed by the crystal structure. Inequivalent ligand environments surrounding the magnetic sites generate direction-dependent hybridization and break inversion or translation symmetry between magnetic sublattices~\cite{T_break_spontaneous_Hall_col_AFM_Smejkal_2020,
research_landscape_of_altermagnets_smejkal_22,
bonding_behind_AM_NiS_Mandal_2025}. 

\begin{figure}
    \centering
    \includegraphics[width=1\linewidth]{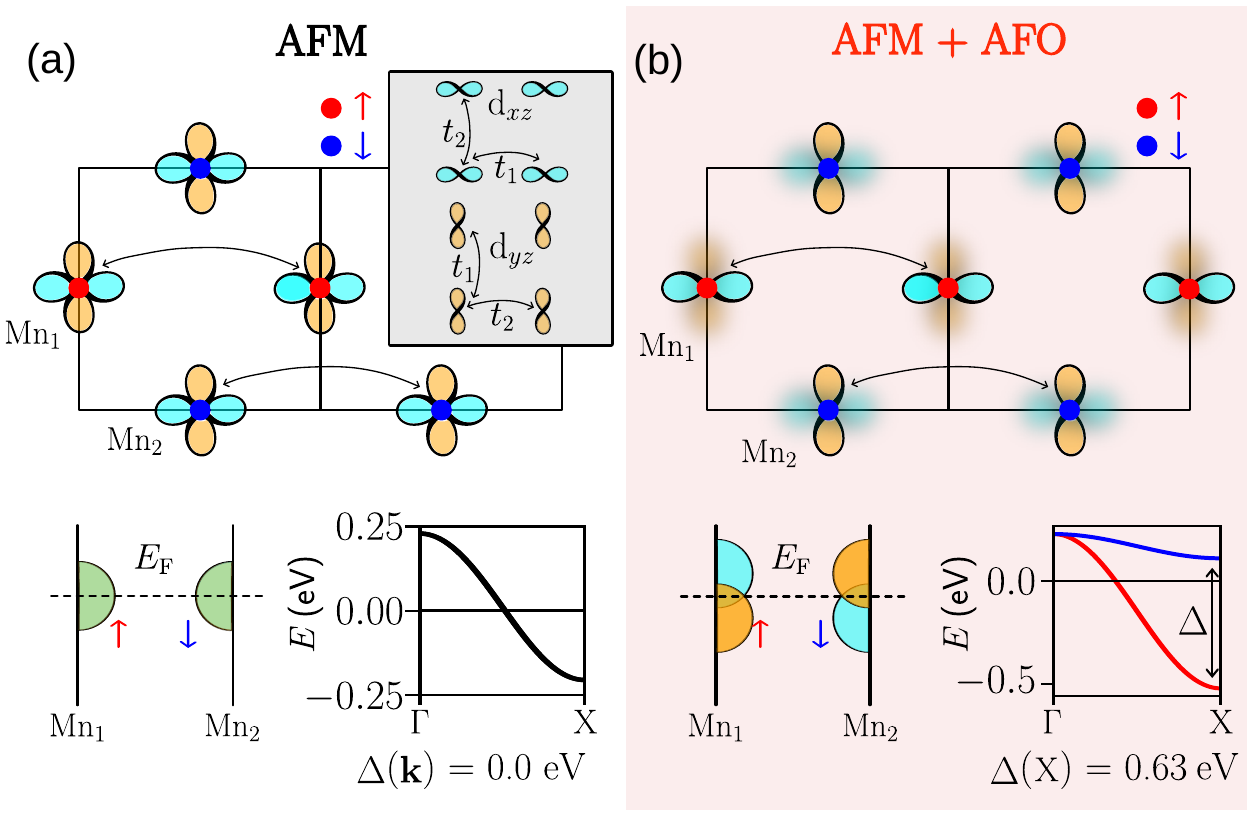}
    \caption{{\bf Spontaneous altermagnetic spin splitting induced by antiferro-orbital  ordering (AFO) in a collinear antiferromagnet (AFM).}
    (a) In the absence of orbital order, sublattice symmetry enforces Kramers-degenerate bands for both spin sectors.
    (b) Antiferro-orbital order breaks the equivalence of magnetic sublattices, leading to orbital-spin locking and a momentum-dependent spin splitting characteristic of altermagnetism.
    \label{fig1}}
\end{figure}

A fundamentally different possibility is that the required anisotropy emerges spontaneously from electronic correlations. In a multi-orbital antiferromagnet, interaction-driven orbital polarization can differentiate the two spin-sublattice sectors even when the lattice itself does not support a ligand-driven altermagnetic phase. Since spin is tied to the sublattice in a collinear antiferromagnet, antiferro-orbital order naturally transfers distinct orbital character to opposite spin channels, generating altermagnetic spin splitting. While theoretical studies have suggested that Fermi-surface instabilities and orbital ordering can realize this scenario~\cite{Fermi_liquid_instabilities_in_spin_Wu_2007,
spontaneous_OO_AM_leeb_2024,
nitin_AM_in_Lieb_lattice_Hubbard_model_25, kaushal_2026_SAM_correlated_system}, no real two-dimensional metallic material has yet been shown to host a correlation-driven altermagnetic state~\cite{layered_metallic_AM_Zhang2025,hydroxyl_rotation_AM_yang_2025,AM_and_strain_induced_AM_Li_2025,strain_induced_AFM_to_AM_atasi_2024,symmetry_breaking_AM_parfenov_2025, search_for_metallic_AM_wan_2025,spin_split_AM_ruo2_chen_2025,TMR_in_AM_sun_2025}. This gap is particularly significant in two dimensions, where electrical control via gating and heterostructure engineering is naturally accessible~\cite{gate_field_control_of_AM_zhang_2024,
switchable_ferroelectric_altermagnets_wang_2025}. Currently, intrinsic monolayer metallic altermagnets with large nonrelativistic spin splitting remain scarce (see Table~\ref{tab:AMs_with_spin_splitting}).

In this Letter, we demonstrate that spontaneous antiferro-orbital order provides a general microscopic route to altermagnetism in correlated two-dimensional metals. We show that monolayer YbMn$_2$Ge$_2$ realizes this mechanism. Its monolayer symmetry permits altermagnetic spin splitting, and correlation-driven d$_{xz}$/d$_{yz}$ orbital order enables microscopic differentiation between opposite spin-sublattice sectors to produce a $d$-wave spin texture with a giant nonrelativistic spin splitting exceeding $1$~eV. The resulting metallic phase supports an exceptionally large transverse spin conductivity with strong gate tunability. These results establish correlation-driven orbital order as a viable route for designing two-dimensional metallic altermagnets. 

\begin{figure*}
    \centering
    \includegraphics[width=1\linewidth]{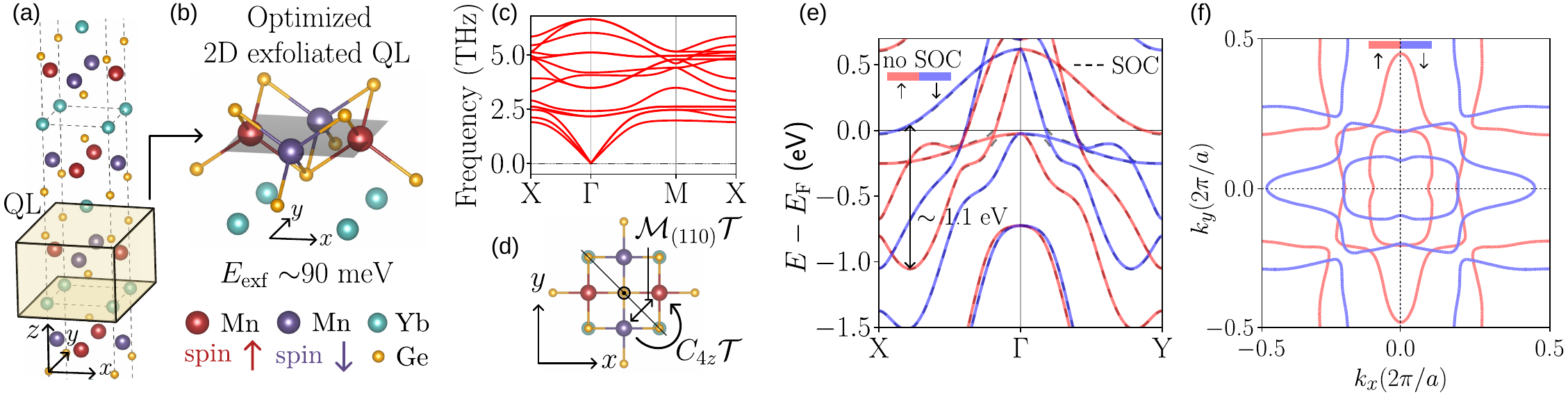}
    \caption{{\bf Monolayer YbMn$_2$Ge$_2$ as a correlated two-dimensional platform for altermagnetism.}
    (a) Bulk crystal structure showing stacking of quadrupole layers (QLs) along the $z$ direction, with adjacent layers related by inversion symmetry.
    (b) Exfoliated two-dimensional unit cell consisting of a single QL, with four Mn atoms forming an $xy$-plane antiferromagnet (red and blue denote opposite spins).
    (c) Phonon spectrum without imaginary modes, confirming dynamical stability of the AFM phase.
    (d) Top view of the monolayer structure. The system preserves $C_{4z}\mathcal{T}$ symmetry and diagonal mirror-time-reversal symmetries, which connect the magnetic sublattices.
    (e) Electronic band structure showing large nonrelativistic spin splitting along the $\mathrm{X}\Gamma\mathrm{Y}$ path, exceeding $1$ eV near the Fermi energy.
    (f) Fermi contours exhibiting a characteristic $d$-wave-like spin-split structure, with red (blue) indicating up-spin (down-spin) polarization.
    }
    \label{fig2}
\end{figure*}
\textit{Orbital-order induced altermagnetism:--}
In a collinear antiferromagnet, altermagnetism emerges when Kramers spin degeneracy is lifted without producing a net magnetization. Microscopically, this requires the two opposite magnetic sublattices to become inequivalent for charge transfer along a given crystallographic direction, so that the effective hopping amplitudes differ for opposite spin channels. In most experimentally realized altermagnets, this anisotropy is imposed directly by the crystal structure through inequivalent ligand environments and direction-dependent hybridization. Here we focus on the complementary case in which the lattice does not itself fix the full microscopic anisotropy, but electronic correlations generate it spontaneously through orbital order.

The essential physics is captured by a collinear bipartite antiferromagnet with two sublattices and anisotropic d$_{xz}$ and d$_{yz}$ orbitals on each site. In the absence of orbital order, the two magnetic sublattices are related by inversion symmetry, and in the collinear AFM state the combined ${\cal P}{\cal T}$ symmetry maps one spin-sublattice sector onto the other. Because the orbital occupations are identical on the two sublattices, the bands remain Kramers degenerate despite opposite local moments. This degeneracy is lifted once spontaneous antiferro-orbital order develops. We characterize the orbital polarization on sublattice $i$ by
\begin{equation}
    \Delta_i = n_i^{xz} - n_i^{yz}~,
\end{equation}
with $\Delta_1 = \Delta_0 = -\Delta_2 \neq 0$, corresponding to $(\pi,\pi)$ orbital order.

Since the d$_{xz}$ and d$_{yz}$ orbitals are spatially anisotropic, their hopping amplitudes along orthogonal directions are different. Denoting the corresponding orbital-resolved dispersions by $E_{xz}(\mathbf{k})$ and $E_{yz}(\mathbf{k})$, the energy of a Bloch state depends on its orbital weight. Since spin is tied to the sublattice in a collinear antiferromagnet, the alternating orbital order makes the spin-up and spin-down carriers acquire different orbital character. The resulting spin splitting is therefore
\begin{equation}
    \Delta(\mathbf{k}) = E_{\uparrow}(\mathbf{k}) - E_{\downarrow}(\mathbf{k})
    = \Delta_0
    \left[
    E_{xz}(\mathbf{k}) - E_{yz}(\mathbf{k})
    \right]~.
\end{equation}
Thus, the spin splitting arises from the interplay between spontaneous orbital polarization and intrinsic orbital-hopping anisotropy. A detailed derivation of $\Delta(\mathbf{k})$ is given in the Appendix.

For the lattice shown in Fig.~\ref{fig1}, with lattice spacing $a$, we have 
\(E_{xz}(\mathbf{k}) = E_0 + t_1 \cos(k_x a) + t_2 \cos(k_y a)\), 
and 
\( E_{yz}(\mathbf{k}) = E_0 + t_2 \cos(k_x a) + t_1 \cos(k_y a)\). 
This  gives the $d$-wave spin splitting
\begin{equation}
    \Delta(\mathbf{k})
    =
    \Delta_0 (t_1 - t_2)
    \big[\cos(k_x a) - \cos(k_y a)\big]~.
\end{equation}
It vanishes along the diagonal directions $k_x = \pm k_y$, reverses sign under $C_{4z}$ rotation, and integrates to zero over the Brillouin zone, consistent with the absence of net magnetization. The $d$-wave structure follows directly from the fourfold lattice symmetry. The magnitude of the spin splitting is controlled by the strength of the orbital polarization.

This establishes spontaneous orbital order in a collinear antiferromagnet as a general microscopic route to altermagnetism. Unlike ligand-driven mechanisms, the decisive anisotropy here is emergent, electronic, and correlation induced. In the following, we show that monolayer YbMn$_2$Ge$_2$ provides a concrete realization of this mechanism in a two-dimensional metal.

\textit{Material realization in monolayer YbMn$_2$Ge$_2$:--}
To demonstrate the correlation-driven orbital-order mechanism in a real two-dimensional metal, we investigate monolayer YbMn$_2$Ge$_2$ (YMG). It is obtained by exfoliating a single quadrupole layer from the bulk I4/mmm structure. Bulk YMG consists of stacked Yb–Ge–Mn–Ge blocks related by inversion symmetry [Fig.~\ref{fig2}(a)]. In the G-type antiferromagnetic ground state, the two Mn layers form a $\mathcal{PT}$-related pair~\cite{jana_YMG_paper_25}. Exfoliation removes this inversion and breaks the interlayer $\mathcal{PT}$ symmetry.

Our calculations show the exfoliation energy of $\sim 90$ meV per unit cell. This is well below the empirical threshold for mechanically isolable 2D materials~\cite{jung_2018,mounet_2018}, indicating experimental feasibility. 
The absence of imaginary phonon modes throughout the whole Brillouin zone confirms the dynamical stability of exfoliated monolayer [Fig.~\ref{fig2}(c)]. 
Within 2D the long range collinear antiferromagnetic order is stabilized by finite in-plane magnetic anisotropy, which circumvents the Mermin-Wagner constraint~\cite{mermin_wagner_1966,Gong2017_CGT_FM, cr2seo_monolayer_AM_Khan2025, cri3_anisotropy_Lado_2017}. The N\'eel state with moments along the easy $y$ axis is lower in energy by $1.1$ meV compared to out-of-plane alignment (see SI). 

In the monolayer geometry [Fig.~\ref{fig2}(b)], the tetrahedral Ge ligands surrounding the two Mn sublattices are rotated by $90^\circ$ relative to each other. As a result, the fractional translation symmetry $t_{1/2}$ present in the bulk is absent. 
Although the two Mn sublattices with their tetrahedral ligand environment seem inversion symmetric, the local charge density distributions are no longer related by inversion symmetry. 
The monolayer therefore satisfies the symmetry condition that allows altermagnetic spin splitting. Our central point, however, is stronger: the large splitting found in YMG is not merely symmetry-allowed, but is microscopically generated and amplified by correlation-driven antiferro-orbital order.

The antiferromagnetic monolayer retains a combined $C_{4z}\mathcal{T}$ symmetry that exchanges the two Mn sublattices [Fig.~\ref{fig2}(d)] and enforces spin degeneracy at the $\Gamma$ point. Away from $\Gamma$, however, the bands split rapidly along both $\Gamma$X and $\Gamma$Y directions, reaching values exceeding 1 eV near the zone boundary [Fig.~\ref{fig2}(e)]. This giant nonrelativistic splitting follows the $d$-wave symmetry structure predicted by the orbital-order mechanism.

The opposite spin splittings along $\Gamma$X and $\Gamma$Y are constrained by diagonal symmetries $\mathcal{M}_{(110)}\mathcal{T}$ and $\mathcal{M}_{(1\bar{1}0)}\mathcal{T}$, which relate $(k_x,k_y)$ to $(k_y,k_x)$ and $(-k_y,-k_x)$, respectively. These symmetries enforce degeneracy along the $\mathrm{M}\Gamma\mathrm{M}$ direction, producing nodal lines in the Fermi contours [Fig.~\ref{fig2}(f)]. Consequently, the spin splitting reverses sign under $90^\circ$ rotation, yielding a symmetry-enforced $d$-wave altermagnetic texture on the Fermi surface.

Monolayer YbMn$_2$Ge$_2$ therefore realizes all key ingredients of a two-dimensional metallic altermagnet: collinear antiferromagnetic order, absence of sublattice-connecting inversion symmetry, and a symmetry-enforced nonrelativistic $d$-wave spin texture with giant ($\sim$1 eV) spin splitting. In the following section, we show that the observed splitting is rooted in correlation-driven antiferro-orbital order, providing a concrete realization of the general mechanism introduced above.

\begin{figure}[htbp]
    \centering
    \includegraphics[width=0.9\linewidth]{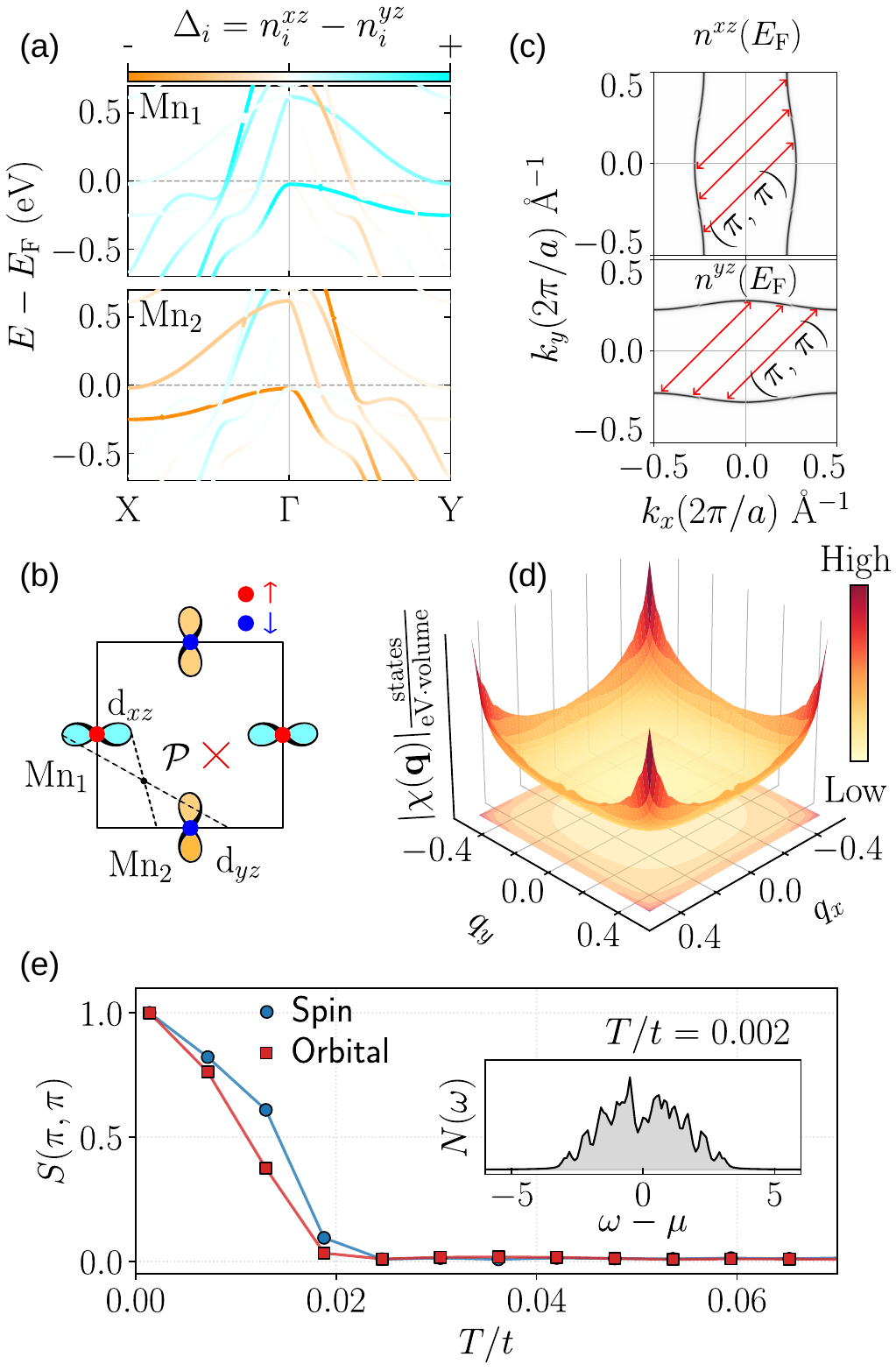}
    \caption{{\bf Correlation-driven orbital order induces altermagnetism in monolayer YbMn$_2$Ge$_2$.}
    (a) Orbital-projected band structure along the X–$\Gamma$–Y path, showing sublattice resolved d$_{xz}$ and d$_{yz}$ orbital polarization from DFT+$U$ calculations.
    (b) Orbital-ordered configuration of the two Mn sublattices: correlation-driven antiferro-orbital order breaks their equivalence, with the spin-up (spin-down) sublattice predominantly occupying the d$_{xz}$ (d$_{yz}$) orbital, producing anisotropic charge distributions in orthogonal planes.
    (c) Fermi contours exhibiting $(\pi,\pi)$ nesting, with dominant d$_{xz}$ (top) and d$_{yz}$ (bottom) orbital character.
    (d) Orbital pseudospin susceptibility showing a pronounced peak at $(\pi,\pi)$, indicating an instability toward antiferro-orbital order.
    (e) Spin and orbital structure factors from semiclassical Monte Carlo simulations of the two-orbital Hubbard model ($U \simeq 1$ eV, $J_H = U/4$, $U' = U - 2J_H$), demonstrating robust $(\pi,\pi)$ antiferromagnetic and antiferro-orbital order. The inset density-of-states plot confirms that the resulting orbital-ordered altermagnetic phase remains metallic.
    \label{fig3}}
\end{figure}
\textit{Correlation-stabilized itinerant altermagnetism:--}
We now demonstrate the emergence of antiferro-orbital polarization in monolayer YbMn$_2$Ge$_2$ and its stabilization by electronic correlations. The site- and orbital-resolved DFT+$U$ band structure along the X–$\Gamma$–Y path [Fig.~\ref{fig3}(a)] shows a clear sublattice-resolved orbital polarization tied to the spin sector: states near the Fermi energy are predominantly d$_{xz}$ on one Mn sublattice and d$_{yz}$ on the other. This staggered polarization is illustrated in Fig.~\ref{fig3}(b), and it corresponds to the $(\pi,\pi)$ antiferro-orbital order introduced in the minimal model.

To understand its microscopic origin, we analyze the orbital-projected Fermi contours of the effective tight-binding model derived from the low-energy DFT bands of Kramers degenerate AFM YMG (see Appendix). As shown in Fig.~\ref{fig3}(c), the d$_{xz}$ and d$_{yz}$ Fermi surfaces exhibit pronounced $(\pi,\pi)$ nesting. This nesting leads to strong peaks at the Brillouin-zone corners in the orbital pseudospin susceptibility [Fig.~\ref{fig3}(d)], indicating an intrinsic instability toward $(\pi,\pi)$ orbital polarization. Here, the pseudospin operator distinguishes d$_{xz}$ and d$_{yz}$ occupations (see Appendix).

To establish that this instability is stabilized by interactions, we consider a two-orbital Hubbard-Kanamori Hamiltonian, 
\begin{equation}
H = H_{\mathrm{TB}} + H_{\mathrm{I}}~.
\end{equation}
$H_{\mathrm{TB}}$ is obtained by downfolding the four low-energy bands near the Fermi level. $H_{\mathrm{I}}$ captures local Coulomb interactions with intra- and inter-orbital repulsion ($U$, $U'$), Hund’s coupling $J_H$, and pair hopping $J' = J_H$ (see SI for details).

We solve this model using semiclassical Monte Carlo simulations~\cite{OSDC_anamitra_2016,monte_carlo_mean_field_1D_Hubbard_anamitra_2014}, which treat interaction-induced auxiliary fields self-consistently while retaining itinerant fermions. For interaction strengths consistent with DFT+$U$ ($U \sim 1$ eV, $J_H = U/4$), the system with $N^2$ sites develops simultaneous $(\pi,\pi)$ antiferromagnetic and antiferro-orbital order. The orbital ordering is quantified by the structure factor
$
S_{\mathrm{orb}}(\mathbf{q}) = N^{-2} \sum_{ij}
\langle \Delta_i \Delta_j \rangle
e^{i\mathbf{q}\cdot(\mathbf{r}_i - \mathbf{r}_j)},
$
with $\Delta_i = n_i^{xz} - n_i^{yz}$ at the Mn lattice site $\textbf{r}_i$. As shown in Fig.~\ref{fig3}(e), both spin and orbital structure factors peak sharply at $(\pi,\pi)$, demonstrating spontaneous symmetry breaking in both sectors. Crucially, the density of states remains finite at the Fermi level (inset), confirming that the ordered phase is metallic.

These results establish that monolayer YbMn$_2$Ge$_2$ realizes a correlation-stabilized itinerant altermagnet, in which long-range antiferromagnetic and antiferro-orbital order coexist in a conducting state. The resulting phase hosts a giant $d$-wave spin splitting that reverses sign under $C_{4z}$ rotation while preserving zero net magnetization. This symmetry-enforced spin texture, arising from orbital-order-driven altermagnetic band structure, directly governs the nonrelativistic spin transport discussed next.

\begin{figure}[t]
    \centering
    \includegraphics[width=1\linewidth]{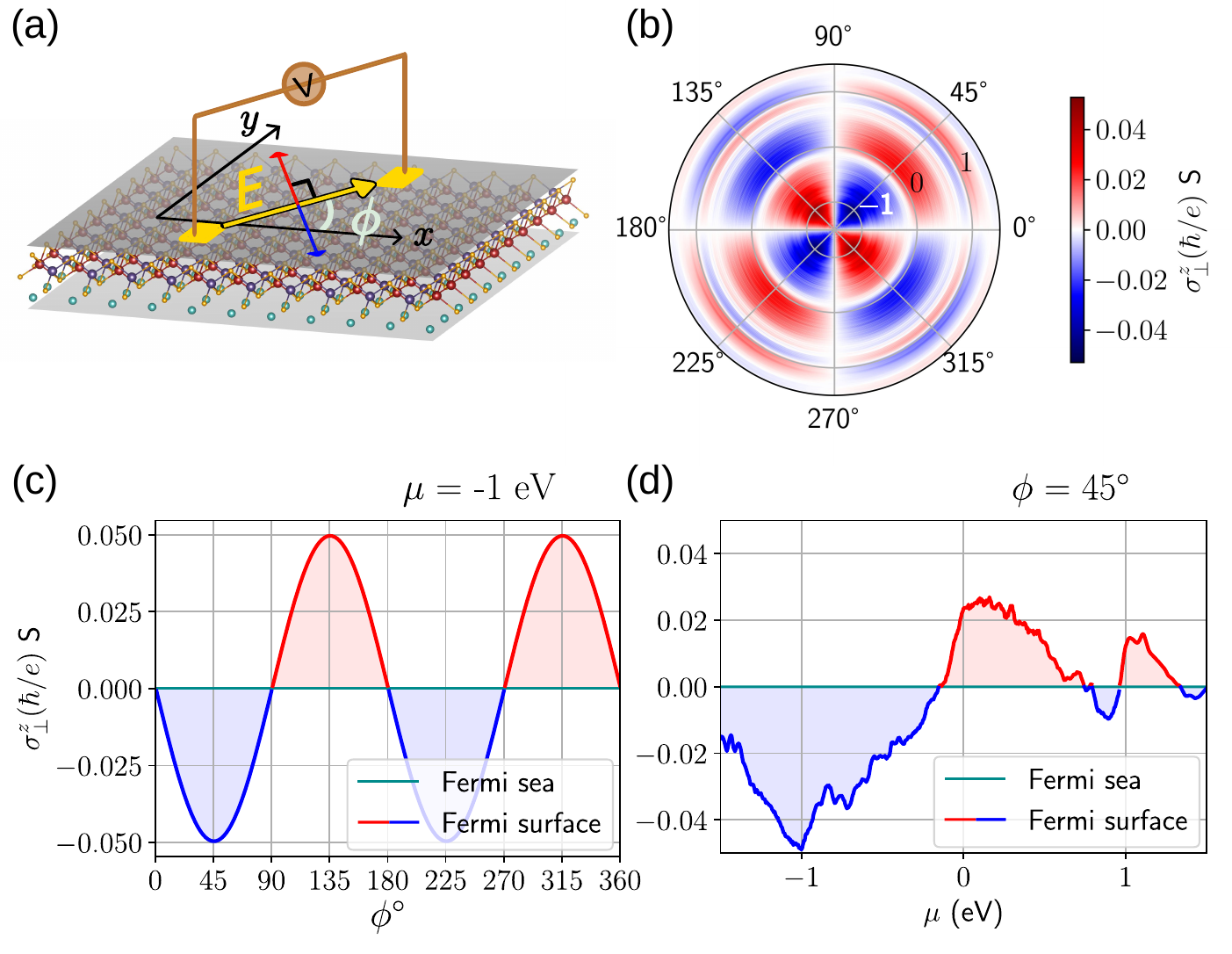}
   \caption{{\bf Giant and anisotropic transverse spin response in monolayer YbMn$_2$Ge$_2$.}
    (a) Schematic device geometry with an in-plane electric field applied at an angle $\phi$ to the $x$ axis. The orbital-order-driven altermagnetic band structure generates a transverse spin current perpendicular to the applied field.
    (b) Fermi-surface contribution to the transverse spin conductivity shown as a polar color map. The radial axis denotes chemical potential $\mu$, and the polar angle represents the direction of the applied electric field. The response has a characteristic $d$-wave angular dependence.
    (c) Angular dependence of the transverse spin conductivity at $\mu = -1$ eV, showing the same $d$-wave symmetry. The Fermi-sea contribution is negligible compared to the Fermi-surface contribution.
    (d) Chemical-potential dependence of the transverse spin conductivity at $\phi = 45^\circ$. The Fermi-surface contribution changes sign near the Fermi energy.
    \label{fig4}}
\end{figure}
\textit{Giant nonrelativistic spin transport:-} 
%
The giant nonrelativistic spin splitting generated by orbital-order-driven altermagnetism naturally leads to a large transverse spin current~\cite{SST_rafael_21,MR_smejkal_22,spin_neutral_currents_for_spintronics_Shao_21}. To quantify this, we calculate both Fermi-surface and Fermi-sea contributions to the spin conductivity. The Fermi-surface contribution for a spin current flowing along $\gamma$ with spin polarization along $\nu$, driven by an electric field along $\lambda$~\cite{gang_su_spin_current_24, sayan_2025_symmetrydrivenintrinsicnonlinearpure, sarkar2025spinbandgeometrydrives}, is given by
\begin{equation}
\label{surf_curr} 
\sigma_{\gamma\lambda}^{\mathrm{surf},\,\nu}
=
e \tau
\sum_m
\int \frac{d\mathbf{k}}{4\pi^2}
\frac{\partial f_m}{\partial \epsilon} 
\,
\langle \psi_m | J_\gamma^{\nu} | \psi_m \rangle
\,
\langle \psi_m | v_\lambda | \psi_m \rangle~.
\end{equation}
Here, the spin-current operator is $ J_\gamma^{\nu} = \frac{1}{2} \{ S^\nu , v_\gamma \}$, with $S^\nu = (\hbar/2)\sigma^\nu$ and $v_\gamma$ the velocity operator, $\tau$ is the momentum relaxation time, and the sum runs over Bloch bands.

In the absence of SOC, the time-reversal-odd transverse spin conductivity $\sigma_{\perp}^{\mathrm{surf},\,z}$ reaches $5\times10^{-2}\,(\hbar/e)\,\mathrm{S}$ [Fig.~\ref{fig4}(a,b)], corresponding to $\sim 5\times10^{5}\,(\hbar/e)\,\mathrm{S/cm}$ when normalized by the monolayer thickness. This value is significantly larger than that reported in thin-film Co$_2$MnGa~\cite{bainsla_2025}, and exceeds that of bulk RuO$_2$~\cite{rafael_2021}. The large magnitude directly reflects the $\sim 1$ eV spin splitting generated by spontaneous antiferromagnetic and antiferro-orbital order.

The angular dependence of $\sigma_{\perp}^{\mathrm{surf},\,z}$ shows a characteristic $d$-wave form,
\begin{equation}
\sigma_{\perp}^{\mathrm{surf},\,z}(\phi)\propto \sin(2\phi)~,
\end{equation}
as shown in Fig.~\ref{fig4}(b,c). This follows directly from the $d$-wave spin texture of the altermagnetic Fermi surface [Fig.~\ref{fig2}(f)]. The Fermi-sea contribution is negligible in the absence of SOC (see SI), indicating that the response is dominated by Fermi-surface quasiparticles.

A key feature is the strong chemical-potential tunability of the transverse spin response. As the Fermi level is varied, $\sigma_{\perp}^{\mathrm{surf},\,z}$ reverses sign [Fig.~\ref{fig4}(b,d)] when the dominant band velocities switch between spin channels, consistent with the spin-resolved dispersions in Fig.~\ref{fig2}(e). This gate-controlled polarity reversal enables direct electrical control of both the sign and magnitude of the spin current, highlighting the potential of two-dimensional altermagnets for spintronic applications.

These results also suggest direct experimental fingerprints of the proposed mechanism. Spin- and momentum-resolved spectroscopy should detect the large $\Gamma$X/$\Gamma$Y spin splitting together with its sign reversal under $90^\circ$ rotation and the nodal degeneracy along $\mathrm{M}\Gamma\mathrm{M}$. In transport, the same $d$-wave symmetry should appear as an angular dependence $\sigma_{\perp}^{z}(\phi)\propto \sin 2\phi$, while electrostatic gating should tune both the magnitude and the sign of the transverse spin response.

\textit{Conclusion:-} 
We show that spontaneous antiferro-orbital order in a collinear antiferromagnetic metal provides a general microscopic route to altermagnetism. In monolayer YbMn$_2$Ge$_2$, the monolayer symmetry allows altermagnetic splitting, and correlations dictate its microscopic form through d$_{xz}$/d$_{yz}$ antiferro-orbital order. This produces a symmetry-enforced $d$-wave spin texture with a giant nonrelativistic spin splitting exceeding 1 eV and a large anisotropic transverse spin response.

More generally, correlation-driven orbital order provides a tunable pathway to engineer altermagnetic phases in two-dimensional metals. This enables control of spin splitting and spin textures via electronic correlations, gating, or strain without relying on relativistic effects. Together with the clear spectroscopic and transport signatures identified here, this makes correlated two-dimensional altermagnets a promising platform for electrically controllable spin functionalities.

\section*{Acknowledgments}
N.J. acknowledges insightful discussions with Soumya Sur on correlated physics, with Sunit Das and Sayan Sarkar on transport-related physics, and with Debasis Dutta on DFT-related aspects. 
A.A. acknowledges funding from the Core Research Grant by the
Anusandhan National Research Foundation (ANRF, Sanction No. CRG/2023/007003), Department of Science and
Technology, India. 
A.C. acknowledges funding by the Deutsche Forschungsgemeinschaft (DFG, German Research Foundation) - TRR 173 - 268565370 (project A03) and TRR 288 - 422213477 (project A09 and B05). 
A.M. acknowledges the use of the NOETHER high-performance cluster at NISER. 
We acknowledge the high-performance computing facility at IIT Kanpur, including HPC 2013, and
Param Sanganak. 

\section{Appendix} 

\subsection{Minimal model for orbital-ordered altermagnetism}

In monolayer YMG, Mn atoms form a checkerboard lattice, while Ge ligands occupy the midpoints between next-nearest-neighbor (nnn) Mn sites (separated by lattice constant $a$) due to the tetrahedral coordination geometry [Fig.~\ref{fig2}(b), (d)]. As a result, ligand-mediated nnn hopping dominates over diagonal nearest-neighbor hopping. This hierarchy is captured by a minimal tight-binding (TB) Hamiltonian extracted from \textit{ab initio} calculations, in which all hopping processes are effectively mapped onto the Mn sites.

\begin{figure}[t]
    \centering
    \includegraphics[width=1\linewidth]{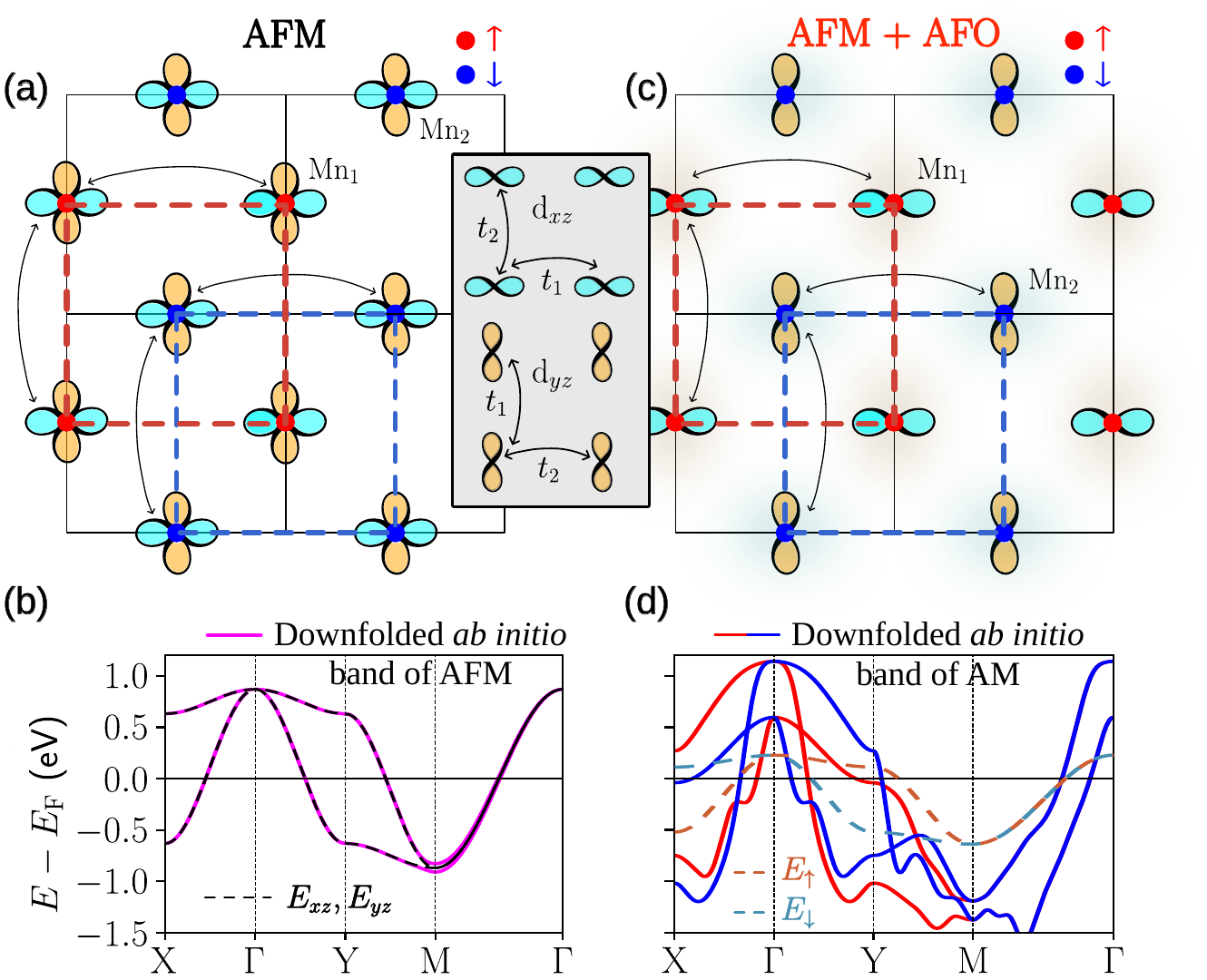}
    \caption{\textbf{Minimal model for orbital-ordered altermagnetism in monolayer YMG.}
    (a) Effective antiferromagnetic lattice with d$_{xz}$ and d$_{yz}$ orbitals on each site. Same-spin sites form two intertwined square lattices, indicated by dashed lines.
    (b) Downfolded AFM band dispersion in the low-energy basis $\{\ket{1,xz,\uparrow}, \ket{1,yz,\uparrow}, \ket{2,xz,\downarrow}, \ket{2,yz,\downarrow}\}$.
    (c) Schematic orbital-ordered configuration.
    (d) Downfolded band dispersion of the altermagnetic state in the same basis, with spin-up and spin-down bands shown in red and blue, respectively.}
    \label{a_fig1}
\end{figure}

Since the dominant hopping is of nnn type, the system can be viewed as two intertwined square lattices (of lattice constant $a$) with effective hopping along the $x$ and $y$ directions [Fig.~\ref{a_fig1}(a)]. The minimal model consists of two orbitals, d$_{xz}$ and d$_{yz}$, on each of the two antiferromagnetically coupled sublattices Mn$_1$ and Mn$_2$. The anisotropic charge distributions of these orbitals generate direction-dependent hopping amplitudes: along $x$, the nnn hopping between d$_{xz}$ (d$_{yz}$) orbitals is $t_1$ ($t_2$), whereas along $y$ the roles of $t_1$ and $t_2$ are interchanged [Fig.~\ref{a_fig1}(a), Fig.~\ref{fig1}(a)].

In the low-energy sector, the dynamics is governed primarily by the majority-spin electrons, such that spin-up (spin-down) electrons predominantly propagate on the Mn$_1$ (Mn$_2$) sublattice. In the collinear antiferromagnetic state, when the orbital composition is identical on the two sublattices, the hopping of the two spin channels remains equivalent along a given direction. This enforces Kramers-degenerate bands, with d$_{xz}$ (d$_{yz}$) states on Mn$_1$ degenerate with those on Mn$_2$.

The effective low-energy Hamiltonian in this Kramers degenerate AFM state, in the basis 
$\{\ket{1,xz,\uparrow}, \ket{1,yz,\uparrow}, \ket{2,xz,\downarrow}, \ket{2,yz,\downarrow}\}$
is
\begin{equation}
H_l =
\begin{pmatrix}
E_{xz} & 0 & 0 & 0 \\
0 & E_{yz} & 0 & 0 \\
0 & 0 & E_{xz} & 0 \\
0 & 0 & 0 & E_{yz}
\end{pmatrix}.
\end{equation}
The orbital-resolved dispersions are given by, 
\begin{align}
E_{xz}(\mathbf{k}) &= E_0 + t_1 \cos(k_x a) + t_2 \cos(k_y a)~, \\
E_{yz}(\mathbf{k}) &= E_0 + t_2 \cos(k_x a) + t_1 \cos(k_y a)~.
\end{align}
Here, $t_1 = 0.375$~eV and $t_2 = 0.059$~eV are extracted from the downfolded low-energy Kramers degenerate antiferromagnetic TB Hamiltonian of YMG. The states $\ket{1,xz,\uparrow}$ and $\ket{2,xz,\downarrow}$, as well as $\ket{1,yz,\uparrow}$ and $\ket{2,yz,\downarrow}$, therefore form Kramers-degenerate antiferromagnetic bands, shown by the dashed black curves in Fig.~\ref{a_fig1}(b). The original downfolded low-energy Kramers degenerate AFM TB model (magenta curve) closely follows these square-lattice dispersions, with only small deviations near the $M$ point due to residual nearest-neighbor hopping.

When orbital order develops [Fig.~\ref{a_fig1}(c)], one sublattice becomes predominantly d$_{xz}$-like while the other becomes d$_{yz}$-like. As a result, the hopping amplitudes for spin-up and spin-down carriers along the $x$ and $y$ directions become unequal, lifting the spin degeneracy. This produces the altermagnetic band splitting observed in the minimal dispersion obtained by downfolding the \textit{ab initio} (DFT + $U$) band structure onto the low-energy AFM basis, as shown in Fig.~\ref{a_fig1}(d). In the next section, we construct a minimal analytical model to explicitly demonstrate how orbital ordering and intrinsic hopping anisotropy jointly generate altermagnetism in an antiferromagnetic metal.

\subsection*{Orbital-order-induced spin splitting}

We now derive the momentum-dependent spin splitting generated by antiferro-orbital order in a collinear bipartite antiferromagnet. We consider a minimal two-sublattice model with anisotropic d$_{xz}$ and d$_{yz}$ orbitals, as shown in Fig.~\ref{fig1}. In the low-energy sector of the collinear AFM state, the relevant states are $\{ \ket{1,\uparrow}, \ket{2,\downarrow} \}$, while $\ket{1,\downarrow}$ and $\ket{2,\uparrow}$ lie at higher energy and are neglected.

The sublattice-projected spin states are written as
\begin{align}
\ket{1,\uparrow}
&=
p_1 \ket{1,xz,\uparrow}
+
q_1 \ket{1,yz,\uparrow}~, \\
\ket{2,\downarrow}
&=
p_2 \ket{2,xz,\downarrow}
+
q_2 \ket{2,yz,\downarrow}~,
\end{align}
where $|p_i|^2 = n_i^{xz}$ and $|q_i|^2 = n_i^{yz}$ denote the orbital weights on sublattice $i$. In this reduced basis, the effective Hamiltonian is diagonal,
\begin{equation}
H_m =
\begin{pmatrix}
E_{\uparrow} & 0 \\
0 & E_{\downarrow}
\end{pmatrix}~,
\end{equation}
with
\begin{align}
E_{\uparrow} &= n_1^{xz} E_{xz}(\mathbf{k}) + n_1^{yz} E_{yz}(\mathbf{k})~, \\
E_{\downarrow} &= n_2^{xz} E_{xz}(\mathbf{k}) + n_2^{yz} E_{yz}(\mathbf{k})~.
\end{align}
The resulting spin splitting is
\begin{equation}
\Delta(\mathbf{k}) =
E_{\uparrow} - E_{\downarrow}
=
\left(n_1^{xz} - n_1^{yz}\right)
\left[E_{xz}(\mathbf{k}) - E_{yz}(\mathbf{k})\right]~.
\end{equation}
This shows that the altermagnetic spin splitting arises from the product of orbital ordering and dispersion anisotropy. Defining the orbital polarization $\Delta_i = n_i^{xz} - n_i^{yz}$ and using $\Delta_1 = \Delta_0 = -\Delta_2$, we obtain the general form
\begin{equation}
\Delta(\mathbf{k})
=
\Delta_0
\left[
E_{xz}(\mathbf{k}) - E_{yz}(\mathbf{k})
\right]~.
\end{equation}
Substituting the lattice dispersions, we obtain
\begin{equation}
\Delta(\mathbf{k})
=
\Delta_0 (t_1 - t_2)
\big[
\cos(k_x a) - \cos(k_y a)
\big]~.
\end{equation}
The spin splitting vanishes along $k_x=\pm k_y$, reverses sign under $C_{4z}$ rotation, and integrates to zero over the Brillouin zone, consistent with the absence of net magnetization.

Using the downfolded Kramers degenerate AFM hopping parameters $t_1 = 0.375$~eV and $t_2 = 0.059$~eV, the model yields a spin-splitting amplitude of $0.632$~eV at X for $\Delta_0 = 1$ [Fig.~\ref{a_fig1}(d)]. This is consistent with the $\sim 0.7$~eV splitting at X obtained from \textit{ab initio} calculations. The full DFT bands with all orbitals exhibit a larger maximum splitting of $\sim 1.1$~eV along the $\Gamma$–X direction, capturing additional effects not included in the minimal square-lattice model. This places monolayer YbMn$_2$Ge$_2$ among the metallic altermagnets with the largest reported spin splitting in 2D (Table~\ref{tab:AMs_with_spin_splitting}).

\begin{table}[t!]
\centering
\begin{tabular}{cccc}
\hline
\hline 
Material & Monolayer~ & ~Metallic & Spin-Splitting (eV) \\
\hline
\hline 
 CrSb \cite{search_for_metallic_AM_wan_2025, band_splitting_in_crsb_Ding_2024}  &  \xmark  &  \cmark  &  0.66 - 0.93  \\
 CrSe \cite{search_for_metallic_AM_wan_2025} &  \xmark  &  \cmark  &  0.47  \\
 {KV$_2$Se$_2$O} \cite{metallic_d_wave_AM_kv2se2o} & \xmark & \cmark & $\sim$1.6 \\
 CaFe4 Al8 \cite{search_for_metallic_AM_wan_2025} &  \xmark  &  \cmark & 0.36  \\
 \hline 
 Cr2SeO \cite{cr2seo_monolayer_AM_Khan2025} &  \shortstack{Janus \\monolayer}  &  \xmark &  0.38 \\
 Fe2Se2O \cite{fe2se2o_monolayer_AM_Wu_2024} &  \cmark  & \xmark  &  0.53 \\
 V$_2$Se$_2$O \cite{v2se2o_monolayer_AM_ma_21} &  \cmark  &  \xmark & $>$ 1  \\
 \shortstack{YbMn$_2$Ge$_2$ \\ (This work)}  &  \cmark  &  \cmark &  $\sim$1.1 \\
\hline
\hline
\end{tabular}
\caption{Representative altermagnets and their characteristic spin-splitting energies.}
\label{tab:AMs_with_spin_splitting}
\end{table}

\subsection{Orbital pseudospin susceptibility}
To quantify the tendency toward orbital ordering and its momentum structure, we calculate the orbital pseudospin susceptibility within linear response theory. The orbital degree of freedom is represented by the pseudospin operator $\tau^z$, which distinguishes the occupations of the d$_{xz}$ and d$_{yz}$ orbitals. A strong enhancement of this susceptibility signals an instability toward orbital order.

The static orbital pseudospin susceptibility is given by \cite{bruus2004many}
\begin{align}
\chi^{zz}_{\mathrm{orb}}(\mathbf{q}) =
-\frac{2}{\mathcal{V}}
\sum_{\mathbf{k},\psi,\phi}
\left| M^z_{\phi\psi}(\mathbf{k},\mathbf{q}) \right|^2
\frac{f(E_{\phi\mathbf{k}})-f(E_{\psi\mathbf{k+q}})}
{i\eta + E_{\phi\mathbf{k}} - E_{\psi\mathbf{k+q}}}~.
\end{align}
Here, $\mathbf{q}$ is the momentum transfer, $\mathcal{V}$ is the system volume, and $\psi,\phi$ label the band indices with eigenvalues $E_{\psi\mathbf{k}}$. The factor of two accounts for spin degeneracy.
The matrix elements
\begin{align}
M^z_{\phi\psi}(\mathbf{k},\mathbf{q})
&=
\langle \phi,\mathbf{k} | \tau^z | \psi,\mathbf{k+q} \rangle
\nonumber\\
&=
\sum_{\alpha\beta}
U^*_{\alpha\phi}(\mathbf{k})
\tau^z_{\alpha\beta}
U_{\beta\psi}(\mathbf{k+q}).
\end{align}
encode the orbital character of the Bloch states. Here, 
$U_{\alpha\psi}(\mathbf{k}) = \langle \alpha | \psi,\mathbf{k} \rangle$
transforms between the orbital basis $\alpha,\beta \in$ \{d$_{xz}$, d$_{yz}\}$ and the band basis, and $\tau^z$ is the Pauli matrix in orbital pseudospin space.

The Fermi-Dirac distribution is denoted by $f(E)$, and $\eta$ is a positive infinitesimal ensuring convergence. Peaks in $\chi^{zz}_{\mathrm{orb}}(\mathbf{q})$ identify the dominant orbital correlations and signal the emergence of antiferro-orbital order at the corresponding wave vector.

\bibliography{references}

\end{document}